\documentclass[a4paper]{jpconf}
\usepackage{graphicx}

\newcommand{\be}{\begin{equation}}
\newcommand{\ee}{\end{equation}}
\newcommand{\bea}{\begin{eqnarray}}
\newcommand{\eea}{\end{eqnarray}}
\newcommand{\bml}{\begin{subequations}}
\newcommand{\eml}{\end{subequations}}
\newcommand{\bfig}{\begin{figure}}
\newcommand{\efig}{\end{figure}}

\begin{document}
\title{Brane inflation: A field theory approach in background supergravity}

\author{Sayantan Choudhury$^{1}$and Supratik Pal$^{1, 2}$}

\address{$^1$Physics and Applied Mathematics Unit, Indian Statistical Institute,
 203 B.T. Road, Kolkata 700 108, India \\$^2$Bethe Center for Theoretical Physics and Physikalisches
Institut der Universit\"{a}t Bonn, Nussallee 12, 53115 Bonn,
Germany}

\ead{sayanphysicsisi@gmail.com, supratik@isical.ac.in}

\begin{abstract}
We propose a model of inflation in the
 framework of brane cosmology driven by background supergravity.
 Starting from bulk supergravity we construct the inflaton potential
 on the brane and employ it to investigate for the consequences to
 inflationary paradigm. To this end, we derive the expressions for
 the important parameters in brane inflation, which are somewhat different
 from their counterparts in standard cosmology, using the one loop radiative
 corrected potential. We further estimate the observable parameters and find
 them to fit well with recent observational data. We have studied extensively
 reheating phenomenology, which explains the thermal history of the universe
 and leptogenesis through the production of thermal gravitino
 pertaining to the particle physics phenomenology of the early universe.
 
\end{abstract}

\section{Introduction}
Investigations for the crucial role of Supergravity  in explaining
cosmological inflation date back to early eighties of the last
century \cite{nills},\cite{Antonio}. One of the generic features of the inflationary paradigm
based on SUGRA is the well-known $\eta$-problem, which appears in
the F-term inflation due to the fact that the energy scale of
F-term inflation is induced by all the couplings via vacuum energy
density.  Precisely, in the expression of F-term inflationary
potential a factor $\exp\left(K/M_{PL}\right)$ appears, leading to
the second slow roll parameter $\eta \gg 1$, thereby violating an
essential condition for slow roll inflation. The usual wayout is
to impose additional symmetry like Nambu-Goldstone shift symmetry or
alternatively to apply Heisenberg symmetry \cite{sayan1}
to solve $\eta$-problem. Another possible way out to smoothen this problem is fine tuning mechanism
via the {\it radion} fields --which decides the separation between visible and invisible brane or so as 
to say brane tension in 
Randall-Sundrum two-brane scenario where
the fifth dimension of 5 D bulk is
compactified on the orbifold $S^{1}/Z_{2}$ of comoving radius R.
This is essential for observationally constraint cosmology on the brane.
 As we will find in the present article the proposed model
of brane inflation  matches quite well with latest observational
data from WMAP \cite{wmap07} and is expected to fit well with
upcoming data from Planck \cite{planck}. Next we have studied in detail reheating
and leptogenesis, which is different
 in the context of brane inflation results in novel features. 
 Hence we have established that this has serious implication for the production of the heavy
   Majorana neutrinos needed for leptogenesis as well as in the 
gravitino phenomenology through analytical and numerical estimations.

\section{Model building from background supergravity}

For systematic development of the formalism, let us 
briefly demonstrate the construction the effective 4D inflationary
potential starting from $N=2, D=5$ SUGRA in
the bulk which leads to an effective $N=1, D=4$ SUGRA in the brane \cite{sayan1}. Considering $S^{1}/Z_{2}$ orbifold compactification in
comoving radius R, the  $N=2,D=5$ bulk SUGRA is described by the following action
\be\begin{array}{ll}\label{su1}
 \displaystyle S=\frac{1}{2}\int d^{4}x\int^{+\pi
R}_{-\pi R}dy\sqrt{g_{5}}\left[M^{3}_{5}\left(R_{(5)}-2\Lambda_{5}\right)+L^{(5)}_{SUGRA}
+\sum^{2}_{i=1}
\delta(y-y_{i})L_{4i}\right].\end{array}\ee Here the orbifold points are 
$y_{i}=(0,\pi R)$ and 
$x^{m}=(x^{\alpha},y)$, where $y$ is bounded in closed interval $[-\pi R,+\pi R]$. Written explicitly, the
contribution from bulk SUGRA in the action
\be\begin{array}{ll}\label{sug2}\displaystyle e^{-1}_{(5)}L^{(5)}_{SUGRA}=-\frac{M^{3}_{5}R^{(5)}}{2}+\frac{i}{2}\bar{\Psi}_{i\tilde{m}}
\Gamma^{\tilde{m}\tilde{n}\tilde{q}}\nabla_{\tilde{n}}
\Psi^{i}_{\tilde{q}}-{S}_{IJ}F^{I}_{\tilde{m}\tilde{n}}F^{I\tilde{m}\tilde{n}}-\frac{1}{2}g_{\alpha\beta}(D_{\tilde{m}}\phi^{\mu})(D^{\tilde{m}}\phi^{\nu})
 + {\rm Fermionic} \\~~~~~~~~~~~~~~~~~~~~~~~~~~~~~~~~~~~~~~~~~~~~~~~~~~~~~~~~~~~~~~~~~~~~~~~~~~~~~~~~~~~~~+ {\rm Chern-Simons},\end{array}\ee
 and including the radion fields ($\chi,T,T^{\dag}$)
 the effective brane SUGRA counterpart turns out
to be $\delta(y)L_{4}=-e_{(5)}\Delta(y)\left[(\partial_{\alpha}\phi)^{\dagger}(\partial^{\alpha}\phi)
+i\bar{\chi}\bar{\sigma}^{\alpha}D_{\alpha}\chi\right]$. Gauging away the Chern-Simons terms 
assuming cubic constraints and $Z_2$ symmetry and appliyng $S^{1}/Z_{2}$
 orbifold compactification the effective 4-dimensional action can be expressed as,
%\sim\frac{1}{2}\Delta(0)P(T,T^{\star})\intd^{4}x\sqrt{g_{4}}\frac{e_{(4)}}{4\pi^{2}R^{2}\sqrt{b_{0}}}\exp\left(\frac{K(\phi,\phi^{\star})}{M^{2}_{PL}}\right)\left[K^{m^{\star}n}\frac{\partial W}{\partial\phi_{m^{\star}}}\frac{\partial W}{\partial\phi_{n}}-3\frac{|W|^{2}}{M^{2}_{PL}}\right]$$$$
\be\begin{array}{ll}\label{ast8}S=\frac{M^{2}_{PL}}{2}\int d^{4}x \sqrt{g_{4}}\left[R_{(4)}+(\partial_{\alpha}\phi^{\mu})^{\dag}(\partial^{\alpha}\phi_{\mu})-QV_{F}-P\int^{+\pi R}_{-\pi R}dy \frac{4(3e^{2\beta y}
+3\lambda^{2}e^{-2\beta y}-2\lambda)}{R^{2}(e^{\beta y}+\lambda e^{-\beta y})^{5}}\right].\end{array}\ee
where the constants are defined in \cite{sayan1}. %and the 4D Planck mass %$M_{PL}=\frac{e_{4}}{b_{0}}=\sqrt{\frac{6}{\lambda}e_{(5)}}=\sqrt{8\pi}M=\frac{M^{3}_{5}}{\sqrt{\lambda}}\sqrt{\frac{3}{4\pi}}=1.22\times 10^{19}GeV$.
  Now using the K$\ddot{a}$hler potential 
$K= \sum_{\alpha}\phi^{\dagger}_{\alpha}\phi^{\alpha}$ and the superpotential
$W=\sum^{\infty}_{n=0}D_{n}W_{n}(\phi^{\alpha})$ with $D_{0}=1$ and $Z_{2}$ symmetry
the renomalizable effective one-loop F-term potential $N=1, D=4$
SUGRA in the brane turns
out to be \cite{sayan1}  
\be\begin{array}{llll}\label{opdsa}
V(\phi)=\Delta^{4}\left[1+\left(D_{4}+K_{4}\ln\left(\frac{\phi}{M}\right)\right)
\left(\frac{\phi}{M}\right)^{4}\right],\end{array}\ee where 
$K_{4}$ and
$D_{4}$ are one loop level constants.
 For our model energy scale $\Delta\simeq 0.2\times10^{16}GeV$
for the window $ -0.70 < D_{4} <-0.60$.
\section{Brane inflation and observational parameters}

 Here we start with Friedmann
equations in brane  
$H^{2}=\frac{8\pi
V}{3M^{2}_{PL}}\left[1+\frac{V}{2\lambda}\right]$. 
 Incorporating the potential of our consideration from
Eq (\ref{opdsa}) the
new  slow roll parameters turn out to be \cite{Maartens}
\be\begin{array}{llllll} \label{first} \epsilon_{V}
=\left(\frac{V^{'}}{V}\right)^{2}\frac{M^{2}_{PL}\left(1+\frac{V}{\lambda}\right)}{16\pi(1+\frac{V}{2\lambda})^{2}},
\eta_{V} =
\left(\frac{V^{''}}{V}\right)\frac{M^{2}_{PL}}{8\pi(1+\frac{V}{2\lambda})},%\\
 \xi_{V} =
\left(\frac{V^{'}V^{'''}}{V^{2}}\right)\frac{M^{4}_{PL}}{(8\pi)^{2}(1+\frac{V}{2\lambda})^{2}},
\sigma_{V} =
\frac{(V^{'})^{2}V^{''''}}{V^{3}}\frac{M^{6}_{PL}}{(8\pi)^{3}(1+\frac{V}{2\lambda})^{3}},\end{array}\ee

Most significantly the generic SUGRA $\eta$-problem is smoothened to some extent in brane cosmology
 due to the presence of $V/\lambda$ term in Friedmann equation. 
The number of e-foldings are  defined in brane cosmology 
for our model as
 \be\begin{array}{llll}\label{noe}
  N\simeq\frac{8\pi}{M^{2}_{PL}}\int^{\phi_{i}}_{\phi_{f}}\left(\frac{V}{V^{'}}\right)\left(1+\frac{V}{2\lambda}\right)d\phi
  \simeq \tiny{\frac{M^{2}}{2U}\left[\left(1+\frac{\alpha}{2}\right)\left(\frac{1}{\phi^{2}_{f}}
-\frac{1}{\phi^{2}_{i}}\right)+\frac{D_{4}}{M^{4}}(1+\alpha)(\phi^{2}_{i}-\phi^{2}_{f})+
\frac{\alpha
D^{2}_{4}}{6M^{8}}(\phi^{6}_{i}-\phi^{6}_{f})\right]},\end{array}
 \ee
where $U=(D_{4}+4K_{4})$. After that the expressions for
amplitude of the scalar and tensor perturbation and
tensor to scalar ratio \cite{Maartens} are given by
 \be\begin{array}{lllllllllllllll}\label{scalar}
\Delta^{2}_{s} \simeq
\frac{512\pi}{75M^{6}_{PL}}\left[\frac{V^{3}}
{(V^{'})^{2}}\left[1+\frac{V}{2\lambda}\right]^{3}\right]_{\star},
%\ee
 \Delta^{2}_{t}  \simeq
\left[\frac{\frac{32}{75M^{4}_{PL}}V\left[1+\frac{V}{2\lambda}\right]}
{\sqrt{1+\frac{2V}{\lambda}\left(1+\frac{V}{2\lambda}
\right)}-\frac{2V}{\lambda}\left(1+\frac{V}{2\lambda}\right)
\sinh^{-1}\left[\frac{1}{\sqrt{\frac{2V}{\lambda}\left(1+\frac{V}{2\lambda}\right)}}\right]}\right]_{\star},%\\
% ~~~~~~~~~~~~~~~~~~~~~~~~~~~~~~~~~~~~~~~~~~~~~~~~~~~~~~
r  =
16\frac{\Delta^{2}_{t}}{\Delta^{2}_{s}}.\end{array}\ee

 Here  $\star$
represents at the horizon
crossing ($k=aH$). Further,  the scale dependence of the perturbations, described by
the scalar and tensor spectral indices and their running as follows
 \be\begin{array}{llllll}\label{tilts}
 n_{s}-1=\frac{d\ln\Delta^{2}_{s}}{d
\ln k}|_{\star}\simeq (2\eta^{\star}_{V}-6\epsilon^{\star}_{V}),~~~~~~~~~~~~~~ n_{t}=\frac{d\ln
\Delta^{2}_{t}}{d\ln k}|_{\star}\simeq -3\epsilon^{\star}_{V},\\
\alpha_{s}=\frac{d\ln n_{s}}{d
\ln k}|_{\star}=(16\eta^{\star}\epsilon^{\star}-18(\epsilon^{\star})^{2}-2\xi^{\star}),~~~~
\alpha_{t}=\frac{d\ln n_{t}}{d
\ln k}|_{\star}=(6\epsilon^{\star}\eta^{\star}-9(\epsilon^{\star})^{2}),
\end{array}\ee
where $d(\ln(k))=Hdt$. Here the consistency conditions
$r=24\epsilon_{V}=24\epsilon^{\star}_{V} ; ~~
n_{t}=-3\epsilon_{V}\simeq-3\epsilon^{\star}_{V}=-\frac{r}{8}$ and
$\frac{d\sigma}{d(\ln(k))}=(\epsilon\sigma-2\eta\sigma)$ are valid in brane cosmology.

\section{Reheating and Leptogenesis in braneworld}
 After inflation when reheating epoch starts 
inflaton decays into different particle constituents. The total inflaton decay width
 participating in leptogenesis can be written as \cite{sayan2}:
\be\begin{array}{llll}\Gamma_{total}(\psi\rightarrow l_{L}{\cal H}, \psi\rightarrow \bar{l}_{L}{\cal H})\simeq \frac{C^2}{16\pi
m_{\phi}}+\frac{h^{2}m_{\psi}}{4\pi}\sim
\frac{1}{(2\pi)^{3}}\left(\frac{\Delta^{6}}{M^{5}}\right)=3H(T^{brh})=\sqrt{\frac{3\rho(t_{reh})}{M^{2}}
\left[1+\frac{\rho(t_{reh})}{2\lambda}\right]}.\end{array}\ee 
Now applying statistical thermodynamics in braneworld the expression for
 transition temperature($T_{c\gamma}$) and the reheating temperature($T^{brh}$)
turns out to be 
\be\begin{array}{lllllll}\label{ubi}T_{c\gamma}
=\sqrt[4]{\left\{C_{\gamma}\frac{\pi^{2}
(K_{4}+4D_{4})^{2}\Delta^{2}_{s}\phi^{6}_{\star}}{\pi^{2}
N^{*}_{\gamma}\alpha^{4}M^{2}}\right\}},~~~~~~~
T^{brh}_{\gamma}=\sqrt[4]{\left\{\frac{W_{\gamma}(K_{4}+4D_{4})\Delta_{s}
\phi^{3}_{\star}\Gamma_{total}}{\pi N^{*}_{\gamma}\alpha^{2}}\right\}}~~~~~~~~~~~~~\forall \gamma\end{array}\ee

where $C_{\gamma}=\left(36000,\frac{288000}{7},19200\right)$, $W_{\gamma}=\left(600,\frac{4800}{7},320\right)$, 
the species index $\gamma=1(B\Rightarrow Boson),2(F\Rightarrow Fermion),
3(M\Rightarrow Mixture)$ and $N^{\star}=N^{\star}_{B}+\frac{7}{8}N^{\star}_{F}$.

\begin{figure}[htb]
% \centerline{\includegraphics[width=7cm, height=5cm]{wmap.eps}}
\centerline{\includegraphics[width=8cm, height=5cm]{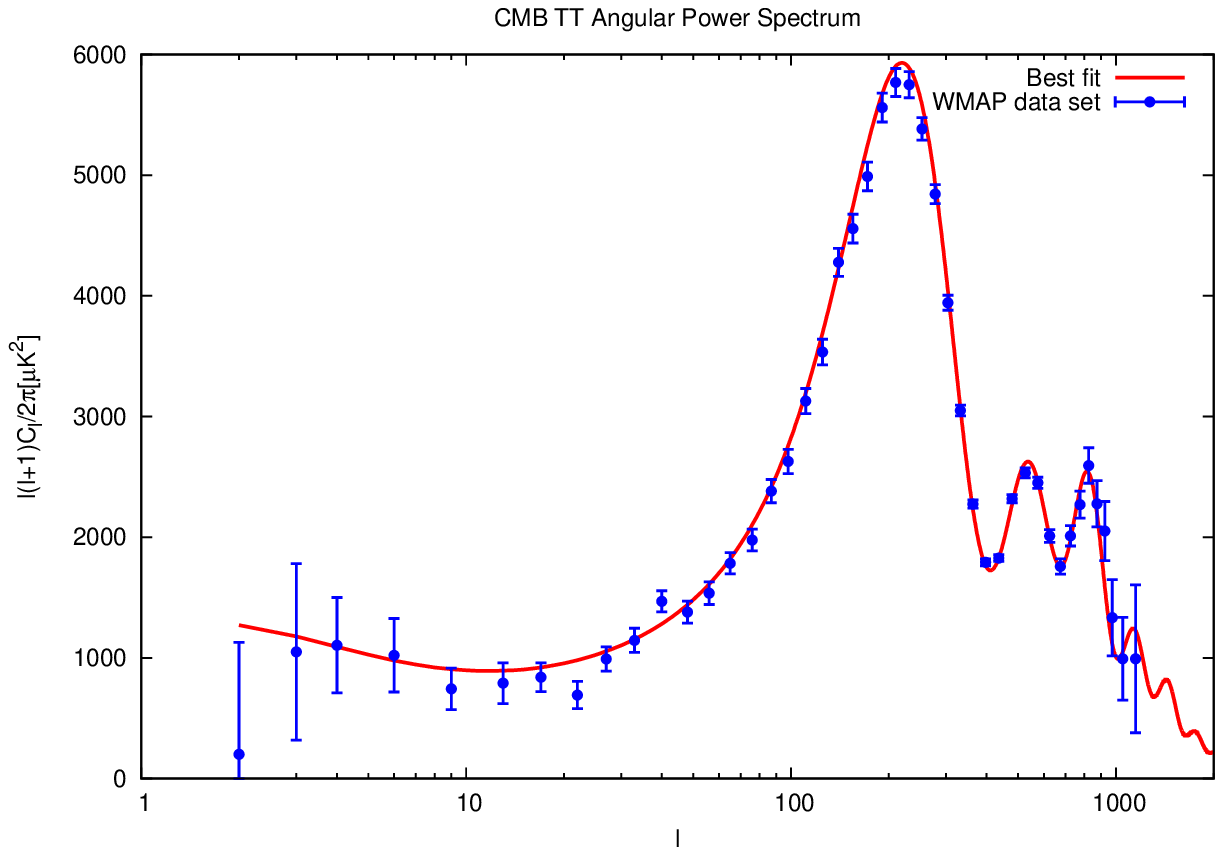} \includegraphics[width=8cm, height=5cm]{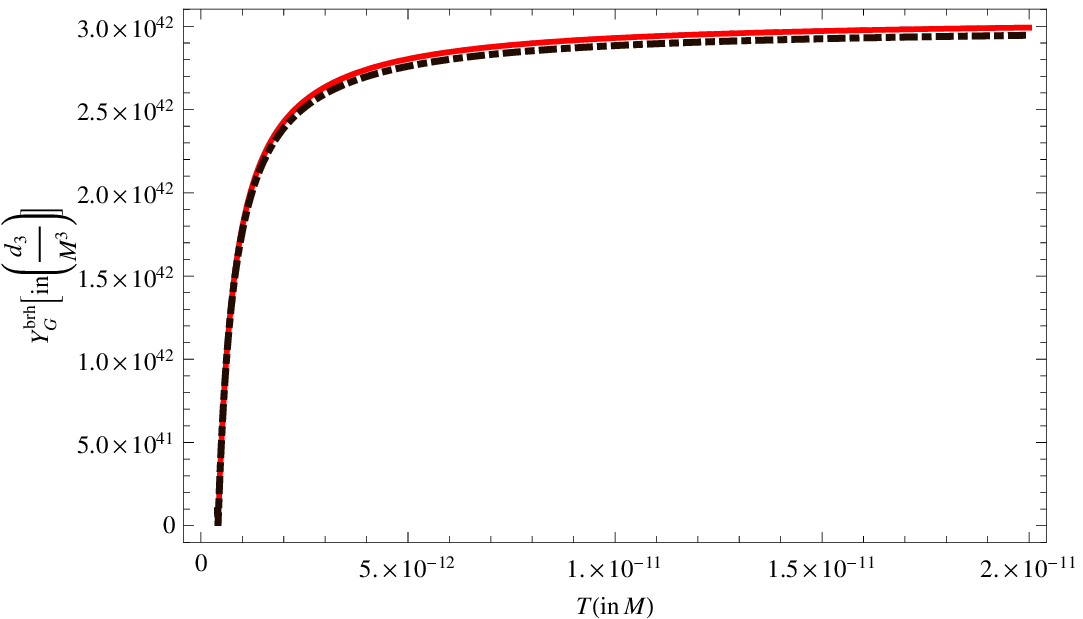}}% \includegraphics[width=6.1cm, height=5cm]{bee.eps}}
 \caption{ Variation of (a) CMB angular power spectrum $C_l^{TT}$
 with the multipoles $l$ for scalar modes, (b)total gravitino abundance vs temperature}\label{tt}
\end{figure}

 To study the impact of $\phi^{4}$ in  
leptogenesis we will start with the Boltzmann equation \cite{turner}
\be\label{rfg} \dot{\rho_{r}}+4H\rho_{r}=\Gamma_{\phi}\rho_{\phi},\ee
where in braneworld high energy limit ($\rho\gg\lambda$) $H^{2}=\frac{4\pi}{3\lambda
       M^{2}_{PL}}\left(\rho_{r}+\rho_{\phi}\right)^{2}$.Now  extremization of the solution of the 
equation(\ref{rfg}) gives extremum temperature during reheating 
%\be\begin{array}{llll}\label{cv1}
 $T^{bh}_{ex}%=\sqrt[4]{\left[\frac{13\sqrt{3}\Delta^{2}M\Gamma_{\phi}}{N^{*}\pi^{2}}\sqrt{\frac{1}{2\alpha}}\right]}
=\sqrt[4]{\left\{\frac{45\Gamma_{\phi}M^{3}_{5}}{8N^{*}\pi^{3}}\right\}}$.%\end{array}\ee
%and it is less than the reheating temperature in brane ($T^{brh}$) which shows devation from GR phenomenology.
 To use this idea in gravitino phenomenology we will start with gravitino Boltzmann equation
  \be\begin{array}{llll}\label{bn1}
\frac{dn_{\tilde{G}}}{dt}+3Hn_{\tilde{G}}=\langle\Sigma_{total}|v|\rangle n^{2}-\frac{m_{\frac{3}{2}}n_{\tilde{G}}}
{\langle E_{\frac{3}{2}}\rangle\tau_{\frac{3}{2}}}.\end{array}\ee
 %where physical maening of each symbols are explicitly mentioned in \cite{sayan2}.
Henve solving equation(\ref{bn1}) using $Y_{\tilde{G}}=\frac{n_{\tilde{G}}}{s}$
 gravitnino abundance at the end of
reheating can be expressed as\cite{sayan2}:
\be\begin{array}{lllll}\tiny\label{opiu}Y^{brh}_{\tilde{G}}(T_{f})=%\left(\frac{}{}\right)
     \left(\frac{45\zeta^{2}(3)\tilde{\alpha}\sqrt{3\lambda}}{2\pi^{7}M\Delta^{2}N^{*}}\right) \left[\left(\frac{60\sqrt{\lambda}}{\pi N^{*}T_{f}}\right)
     \left(1-\frac{T_{f}}{T^{brh}}\right)+\left(\frac{\pi(T^{bh}_{ex})^{4}}{32\Delta^{2}T^{brh}}\right)
\left(32\left(\frac{T^{brh}}{T^{bh}_{ex}}\right)^{4}-1\right)^{\frac{1}{4}}\right.\\ \left.~~~~~~~~~~~~~~~~~~~~~~~~~~~~~~~~~~~~~~~~~~~~~~~~~~~~~~~~~~~~~~~~~~~~~~~~~~~~~~~~~~
 \times\left\{\frac{1}{\Gamma(\frac{3}{4})}+\frac{1}{\Gamma(-\frac{5}{4})}-\frac{2}{\Gamma(-\frac{1}{4})}\right\}\right].\end{array}\ee

which is plotted with respect to temperature in figure(\ref{tt}(b)).

\section{Numerical estimations and data analysis}

 For a typical value of $C_{4}
\simeq D_{4}=-0.7$, $56<N<70$ and decay width
 $\Gamma_{\phi}\simeq2.9\times 10^{-3}GeV$,  mass of the
inflaton $m_{\phi}\simeq10^{13}GeV$ we have: $\Delta^{2}_{s}\sim (1.440-3.126)\times 10^{-9} $,
$\Delta^{2}_{s}\sim 10^{-14} $, $n_{s}\sim(0.936-0.951) $, $n_{t}\sim 10^{-5}$, $r\sim(2.176-4.723)\times10^{-5}$, 
$\alpha_{s}\sim-10^{-3}$, $\alpha_{t}\sim 10^{-6}$, for boson $T^{brh}_{B}\simeq7.6\times 10^{10}GeV$, $T_{cB}\simeq3.2\times 10^{14}GeV$
 for fermion 
$T^{brh}_{F}\simeq7.8\times 10^{10}GeV$, $T_{cF}\simeq3.3\times 10^{14}GeV$ and
 for mixture of species $T^{brh}_{M}\simeq6.5\times 10^{10}GeV$, $T_{cM}\simeq2.8\times 10^{14}GeV$,
 $ T^{bh}_{ex}\simeq7.0\times10^{10}GeV$, final temperature
 $T_{f}\simeq10^{6}GeV$, at the end of reheating $Y^{b-rad}_{\tilde{G}}(T_{f})\simeq2.1\times 10^{-13}GeV^{-3}d_{3}$, where
 %We have calculated all the abundances in the fundamental unit of
 $d_{3}=6.594\tilde{\alpha}\times10^{11}GeV^{3}$ and $\tilde{\alpha}$ is a dimensionless
constant in MSSM. Now using CAMB \cite{camb} with 
the pivot scale $k_0=0.002~{\rm Mpc}^{-1}$, $H_0\sim 71.0 km/sec/MPc$,
 $\tau_{Reion}\sim 0.09$, $\Omega_b h^2\sim 0.022$, $\Omega_c h^2\sim0.111$ and $T_{CMB}\sim2.725K$ we get
$t_0\sim13.707Gyr$, $z_{Reion}\sim10.704$, $\Omega_m\sim 0.267$, $\Omega_{\Lambda}\sim0.732$, $\Omega_k\sim0$,
$\eta_{Rec}\sim 285.10$ and $\eta_0\sim14345.1$.

\section{Summury}
In this article we have proposed a model of brane inflation derived from $N=2, D=5$ supergravity in
the bulk leads to an effective $N=1, D=4$ supergravity in the brane including one-loop radiative corrections.
Then the model has been employed in
estimating the observable parameters, both analytically and
numerically, leading to more precise estimation of the quantities
using the publicly available code CAMB \cite{camb}, which
reveals consistency of our model with latest observations. 
Hence we have studied reheating and leptogenesis in brane by analyzing
 the reheating temperature, followed
by analytical and numerical estimation of different phenomenological parameters showing deviations from GR results.

\section*{Acknowledgments}

SC thanks Council of Scientific and
Industrial Research, India for financial support through Senior
Research Fellowship (Grant No. 09/093(0132)/2010).

\end{document}